\documentclass[aps,floatfix,superscriptaddress,twocolumn,letterpaper]{revtex4}

\usepackage{graphicx}

\newcommand{\lbl}{\affiliation{Nuclear Sciences Division, Lawrence Berkeley National Laboratory, Berkeley, California 94720, USA}}
\newcommand{\lblesd}{\affiliation{Earth Sciences Division, Lawrence Berkeley National Laboratory, Berkeley, California 94720, USA}}
\newcommand{\berkeley}{\affiliation{Physics Department, University of California at Berkeley, Berkeley, California 94720, USA}}
\newcommand{\stanford}{\affiliation{Department of Physics, Stanford University, Stanford, California 94305, USA}}

\newcommand{\nuebar}       {\bar\nu_{\rm e}}

\newcommand{\A}[2]         {$^{#2}$#1}

\begin{document}

\title{A geoneutrino experiment at Homestake}

\author{N.~Tolich}\lbl
\author{Y.-D.~Chan}\lbl
\author{C.A.~Currat}\lbl
\author{M.P.~Decowski}\berkeley
\author{B.K.~Fujikawa}\lbl
\author{R.~Henning}\lbl
\author{K.T.~Lesko}\lbl
\author{A.W.P.~Poon}\lbl
\author{K.~Tolich}\stanford
\author{J.~Wang}\lblesd

\date{\today}

\begin{abstract}
A significant fraction of the 44TW of heat dissipation from the Earth's interior is believed to originate from the decays of terrestrial uranium and thorium. The only estimates of this radiogenic heat, which is the driving force for mantle convection, come from Earth models based on meteorites, and have large systematic errors. The detection of electron antineutrinos produced by these uranium and thorium decays would allow a more direct measure of the total uranium and thorium content, and hence radiogenic heat production in the Earth. We discuss the prospect of building an electron antineutrino detector approximately $700\,{\rm m^3}$ in size in the Homestake mine at the 4850' level. This would allow us to make a measurement of the total uranium and thorium content with a statistical error less than the systematic error from our current knowledge of neutrino oscillation parameters. It would also allow us to test the hypothesis of a naturally occurring nuclear reactor at the center of the Earth.
\end{abstract}

\maketitle

\section{Introduction}
\label{intro}
Thanks in part to Ray Davis' pioneering neutrino experiment\cite{Davis:1968} located in the Homestake mine(44.35$^\circ$\,N, 103.75$^\circ$\,W), more is now known about the interior workings of the Sun than the Earth. The KamLAND collaboration has recently investigated electron antineutrinos originating from the interior of the Earth\cite{Araki:2005}; however, the sensitivity achieved was limited by a large background from surrounding nuclear power reactors. A similar experiment located deep underground to reduce cosmic-ray backgrounds, and away from nuclear power plants, could reach a sensitivity that would allow constraints to be placed on our current knowledge of the Earth's interior.

The idea of using electron antineutrinos, $\nuebar$'s, to study processes inside the Earth was first suggested by Eder\cite{Eder:1966} and Marx\cite{Marx:1969}. \A{U}{238}, \A{Th}{232}, and \A{K}{40} decays within the Earth are believed to be responsible for the majority of the current radiogenic heat production, which is the driving force for Earth mantle convection, the process which causes plate tectonics and earthquakes. These decays also produce $\nuebar$'s, the vast majority of which reach the Earth's surface since neutrinos hardly interact with matter, allowing a direct measurement of the total Earth radiogenic heat production by these isotopes.

The regional composition of the Earth is determined by a number of different methods. The deepest hole ever dug penetrates $12\,{\rm km}$ of the crust\cite{Kremenetsky:1986}, allowing direct sampling from only a small fraction of the Earth. Lava flows bring xenoliths, foreign crystals in igneous rock, from the upper mantle to the surface. The regional composition of the Earth can also be modeled by comparing physical properties determined from seismic data to laboratory measurements. Our current knowledge suggests that the crust and mantle are composed mainly of silica, with the crust enriched in U, Th, and K. The core is composed mainly of Fe but includes a small fraction of lighter elements. Table~\ref{tab:radioactiveConc} shows the estimated concentration of U, Th, and K in the different Earth regions.

Models of Earth composition based on the solar abundance data\cite{McDonough:1995} establish the composition of the undifferentiated mantle in the early formation stage of the Earth, referred to as ``Bulk Silicate Earth" (BSE). Table~\ref{tab:radioactiveConc} includes the estimated concentration of U, Th, and K in the BSE model. The ratio of Th/U by weight, between 3.7 and 4.1\cite{Rocholl:1993}, is known better than the total abundance of each element.

The rate of radiogenic heat released from U, Th, and K decays are $98.1\,{\rm \mu W\,kg^{-1}}$, $26.4\,{\rm\mu W\,kg^{-1}}$, and $0.0035\,{\rm\mu W\,kg^{-1}}$\cite{Schubert}, respectively. Table~\ref{tab:radiogenicHeat} summarizes the total radiogenic heat production rate of these elements in the Earth regions based on the masses and concentrations given in Table~\ref{tab:radioactiveConc}. For comparison, the rate of mantle heating due to lunar tides is a negligible $\sim0.12\,{\rm TW}$\cite{Zschau:1986}.

\begin{table}[t]
\caption{Estimated total mass of the major Earth regions\cite{Schubert}, and the estimated concentration of U, Th, and K in each region. It is assumed that there is no U, Th, or K in the Earth's core. The concentration of radiogenic elements in the mantle is obtained by subtracting the isotope mass in the crust from the Bulk Silicate Earth (BSE) model.}
\centering
\label{tab:radioactiveConc}
\begin{tabular}{lrrrr}
\hline\hline
Region&Total mass&\multicolumn{3}{c}{Concentration}\\[3pt]
&[$10^{21}\,{\rm kg}$]&U[ppb]&Th[ppb]&K[ppm]\\[3pt]
\hline
Oceanic crust\cite{Taylor}&6&100&220&1250\\
Continental crust\cite{Rudnick:1995}&19&1400&5600&15600\\
Mantle&3985&13.6&53.0&165\\
BSE\cite{McDonough:1995}&4010&20.3&79.5&240\\
\hline\hline
\end{tabular}
\end{table}

\begin{table}[t]
\caption{Radiogenic heat production rate in different Earth regions.}
\centering
\label{tab:radiogenicHeat}
\begin{tabular}{lrrrr}
\hline\hline
Region&U&Th&K&Total\\[3pt]
&[TW]&[TW]&[TW]&[TW]\\[3pt]
\hline
Oceanic crust&0.06&0.03&0.03&0.12\\
Continental crust&2.61&2.81&1.04&6.46\\
Mantle&5.32&5.57&2.30&13.19\\
BSE&7.99&8.42&3.37&19.78\\
\hline\hline
\end{tabular}
\end{table}

The radiogenic heat production within the Earth can be compared to the measured heat dissipation rate  at the surface. Based on the rock conductivity and temperature gradient in bore holes measured at 20,201 sites, the estimated heat dissipation rate from oceanic and continental crust, respectively, is $31.2\pm0.7\,{\rm TW}$ and $13.0\pm0.3\,{\rm TW}$, resulting in a total of $44.2\pm1.0\,{\rm TW}$\cite{pollack:1993}. In this study the majority of the heat is lost through the oceanic crust, despite the fact that the continental crust contains the majority of the radiogenic heat producing elements. A recent re-evalutaion of the same data\cite{Hofmeister:2005} suggests that the heat dissipation rate in the oceanic crust is significantly less, resulting in a total heat dissipation rate of $31.0\pm1.0{\rm \,TW}$. The measured heat flow per unit area at the Earth's surface surrounding the Homestake mine\cite{HeatFlow:2005} is consistent with the continental crust average, which suggests that increased local uranium concentration is not significant. 

The Urey ratio, the ratio between mantle heat dissipation and production, indicates what fraction of the current heat flow is due to primordial heat. Subtracting the continental crust heat production rate of 6.5\,TW, the mantle is dissipating heat at a rate of 37.7\,TW and, assuming the BSE model, generating heat at a rate of 13.2\,TW, giving a Urey ratio of $\sim$0.35. It is widely believed that the mantle convects although the exact nature of that convection is still unclear. Models of mantle convection give Urey ratios greater than $\sim$0.69\cite{Richter:1984,Jackson:1984,Spohn:1982}, which is inconsistent with the value obtained from heat considerations. A direct measurement of the terrestrial radiogenic heat production rate would help our understanding of this apparent inconsistency.

\section{Geoneutrino signal}
\label{sec:geoneutrinos}
\begin{figure}
\centering
\includegraphics[width=0.9\columnwidth]{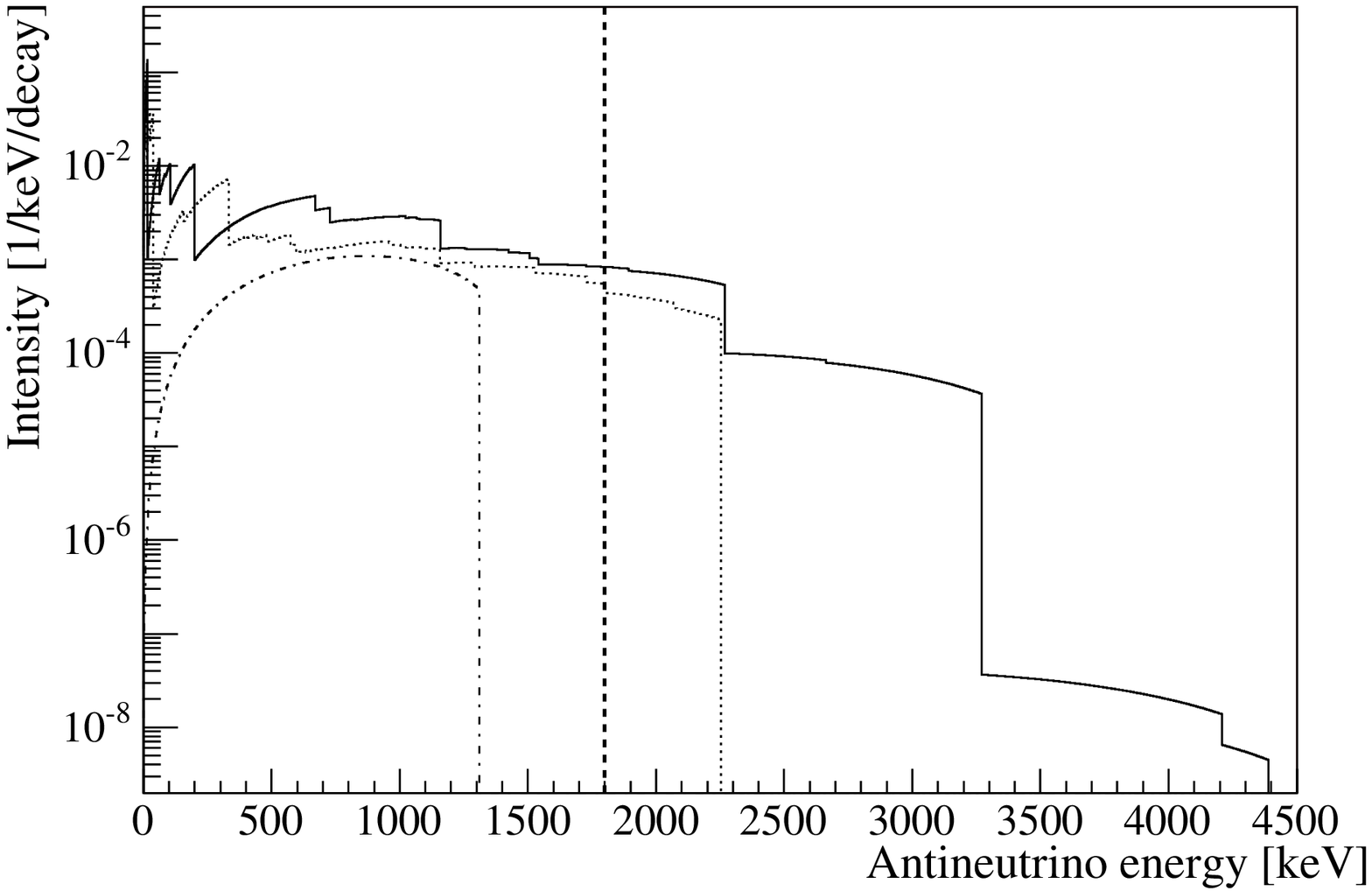}
\caption{The $\nuebar$ energy distributions for the \A{U}{238} (solid), \A{Th}{232}(dash), and \A{K}{40}(dot-dash) decay chains. The vertical line represents the $\nuebar$ detection threshold for neutron inverse $\beta$ decay. Only the \A{U}{238} and \A{Th}{232} chains are measurable with neutron inverse $\beta$ decay.}
\label{fig:spectrums} 
\end{figure}

A $\nuebar$ is produced whenever a nucleus $\beta^-$ decays. The \A{U}{238} and \A{Th}{232} decay chains\cite{TOI} both contain at least four $\beta^-$ decays, and \A{K}{40} $\beta^-$ decays with a branching fraction of 89.28\%. These $\beta^-$ decays result in the well established $\nuebar$ energy distributions for \A{U}{238}, \A{Th}{232}, and \A{K}{40} shown in Figure~\ref{fig:spectrums}. Because $\nuebar$'s have such a small cross-section for interaction with matter, the majority of these $\nuebar$'s produced within the Earth reach the surface. However, due to a phenomenon usually referred to as ``neutrino oscillation", the $\nuebar$ may change into a ${\bar\nu_{\rm \mu}}$ or ${\bar\nu_{\rm \tau}}$. The probability of the $\nuebar$ being found in the same state as a function of distance traveled, $L$, can be approximated as,
\begin{equation}
P(E_\nu ,L) = 1-\sin^22\theta_{12}\sin^2\left(\frac{1.27\Delta m^2_{12}[{\rm eV^2}]L[{\rm m}]}{E_\nu[{\rm MeV}]} \right),
\end{equation}
where $\Delta m^2_{12}=7.9^{+0.6}_{-0.5}\times10^{-5}\,{\rm eV}^2$, and $\sin^22\theta_{12}=0.816^{+0.073}_{-0.070}$\cite{Araki:2004}. This assumes two ``flavor" oscillation and neglects ``matter effects" both of which are less than 5\,\% corrections\cite{Araki:2005}.

The most common method\cite{Apollonio:2003,Boehm:2001,Araki:2005} for detecting $\nuebar$'s is neutron inverse $\beta$ decay,
\begin{equation}
\nuebar+{\rm p}\longrightarrow {\rm e}^++{\rm n}.
\end{equation}
The detection of both the positron, e$^+$, and neutron, n, separated by a small distance and time, greatly reduces the number of backgrounds. Due to the reaction threshold, the minimum $\nuebar$ energy detectable by this method is 1.8\,MeV, which has the disadvantage that the \A{K}{40} $\nuebar$'s cannot be detected since they have a maximum energy of 1.3\,MeV. To zeroth order in 1/M, where M is the nucleon mass, the total positron energy, $W^{(0)}_e$, is related to the total antineutrino energy, $W_\nu$, by
\begin{equation}
W^{(0)}_e = W_\nu - m_n + m_p,
\end{equation}
where $m_n$ and $m_p$ are the neutron and proton masses, respectively. Therefore, the $\nuebar$ energy can be estimated from a measurement of the positron kinetic energy. This allows spectral separation of the $\nuebar$'s from \A{U}{238} and \A{Th}{232} decays.

\begin{figure}
\centering
\includegraphics[width=0.9\columnwidth]{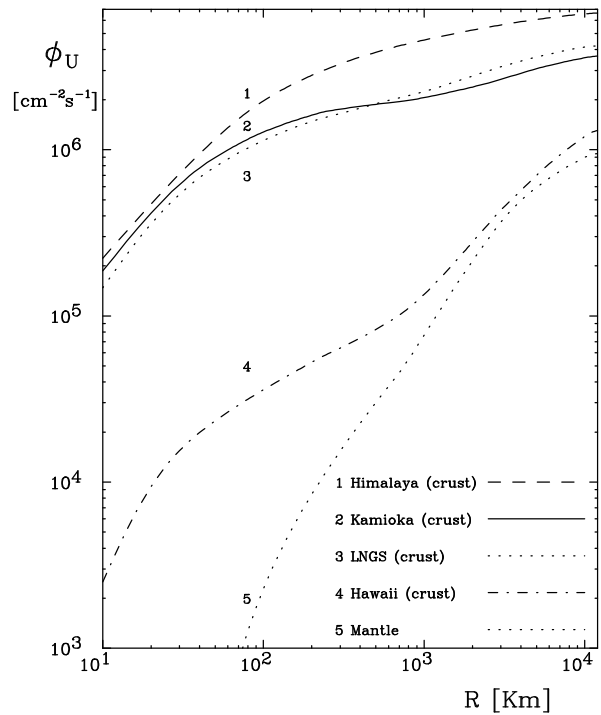}
\caption{Cumulative geoneutrino flux as a function of distance to the source\cite{Mantovani:2004}. The Himalaya curve is for a detector located over thick continental crust. The Kamioka curve is for a detector located at the boundary of continental and oceanic crust. The LNGS curve is for a detector located over continental crust; this probably best represents a detector located at the Homestake mine. The Hawaii curve is for a detector located over oceanic crust; this most closely matches the flux from the mantle, since there is no high local uranium and thorium concentrations.}
\label{fig:fluxDistance} 
\end{figure}

The geoneutrino observation rate depends on the decay rate of \A{U}{238} and \A{Th}{232}, the resulting $\nuebar$ energy distribution, the detection cross-section, the neutrino oscillation parameters, and the distribution of the \A{U}{238} and \A{Th}{232} in the Earth. Based on a detailed simulation\cite{Mantovani:2004}, including seismic models of crustal thickness, the number of neutron inverse $\beta$ decays at Homestake due to terrestrial \A{U}{238} and \A{Th}{232} is estimated to be $54$ per $10^{32}$ target protons per year, assuming $\sin^22\theta_{12}=0.816$. The lines labeled 1, 2, and 3 in Figure~\ref{fig:fluxDistance} show the cumulative geoneutrino fluxes as a function of distance from detectors located over continental crust of varying thickness and with varying contributions from neighboring oceanic crust. A detector located in the Homestake mine could expect $\sim 50\,\%$ of the geoneutrino flux originating within $\sim 500\,{\rm km}$ of the detector.

With $\sim 50\,\%$ of the geoneutrino flux originating within $\sim 500\,{\rm km}$ of the detector it is important to remove the effects of local geology to obtain a global measurement of the total \A{U}{238} and \A{Th}{232} concentration. The estimated error in the signal from local geology for the recent KamLAND geoneutrino measurement is 16\,\%\cite{Sanshiro:2005}. The nearest known uranium reserve\cite{EIA:2004} is located $\sim 100\,{\rm km}$ from the Homestake mine at the boundary of Wyoming and South Dakota. To place an upper limit on the impact of local concentrations of uranium and thorium, we assume that the Earth's total reasonably assured uranium reserves of 3600\,kton uranium\cite{WNA:2006} were located 100\,km from the proposed detector. This would contribute less than $0.03\,\%$ to the expected global signal. A possible heat flow measurement in the Homestake mine and uranium and thorium concentrations obtained from the Homestake mine core samples\cite{SDGS:2006} could be used to reduce the systematic uncertainties associated with geoneutrinos originating from within $\sim 10\,{\rm km}$ of the detector.

It has been suggested that a large amount of uranium may be located in the core of the Earth\cite{Herndon:2003} forming a natural nuclear reactor. This could produce up to $6\,{\rm TW}$ of heat, powering the Earth's dynamo. It would also lead to \A{He}{3} production which could explain the observed anomaly in the $^3{\rm He}/^4{\rm He}$ ratio for gases from the Earth's mantle. Excluding neutrino oscillation, a natural reactor at the Earth's core would produce a very similar energy spectrum of detected  $\nuebar$'s to that from commercial nuclear power reactors, which is peaked at $\sim 4\,{\rm MeV}$ and extends up to $\sim 9\,{\rm MeV}$, see Figure~\ref{fig:spectrumsObserved}. In order to accurately test this hypothesis it is necessary to have a very low commercial nuclear reactor background. The number of $\nuebar$'s detected by neutron inverse $\beta$ decay near the surface of the Earth is expected to be approximately $40$ per $10^{32}$ target protons per year due to a 6\,TW nuclear reactor at the Earth's core, assuming $\sin^22\theta_{12}=0.816$.

\section{Backgrounds}
\label{sec:backgrounds}
The detection of correlated signals in neutron inverse $\beta$ decay significantly enhances the detectability of the geoneutrinos. Nevertheless, other events contribute backgrounds to the measurement. Backgrounds can typically be subdivided into three main categories: natural radioactivity, cosmic-rays and associated spallation products, and other $\nuebar$ sources. The most significant backgrounds in the recent KamLAND geoneutrino measurement\cite{Araki:2005} were $\nuebar$'s from nearby nuclear power reactors and $^{13}{\rm C}(\alpha,{\rm n})$ reactions where the $\alpha$ is primarily from \A{Pb}{210} decay.

\subsection{$\nuebar$ sources}
\label{sec:neutrinoBackgrounds}
\begin{figure}
\centering
\includegraphics[width=0.9\columnwidth]{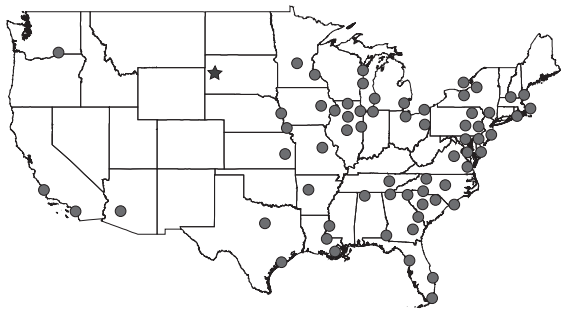}
\caption{Location of nuclear power reactors(circles) in the USA, modified from Ref.\cite{INSC:2005}. The closest nuclear power reactor to Homestake(star) is $\sim 750\,{\rm km}$ away.}
\label{fig:reactors} 
\end{figure}

Figure~\ref{fig:reactors} shows that the Homestake mine is located more than 750\,km away from any major nuclear power reactor. Based on the rated maximum thermal power, and excluding neutrino oscillation, the expected rate of $\nuebar$'s from nuclear reactors is calculated to be 64 per $10^{32}$ target proton yr. Since the $\nuebar$'s typically travel distances greater than 1000\,km, the $\nuebar$ survival probability due to neutrino oscillation can be approximated by $P(E_\nu ,L)\approx 1-0.5\sin^22\theta_{12}$, which equals 0.592 assuming $\sin^22\theta_{12}=0.816$. Therefore, the expected rate in the geoneutrino region, below 3.4\,MeV, and including neutrino oscillation, is only 11 per $10^{32}$ target proton yr, which is $\sim 7\,\%$ of the expected rate at KamLAND\cite{Araki:2005}.

\begin{figure}
\centering
\includegraphics[width=0.9\columnwidth]{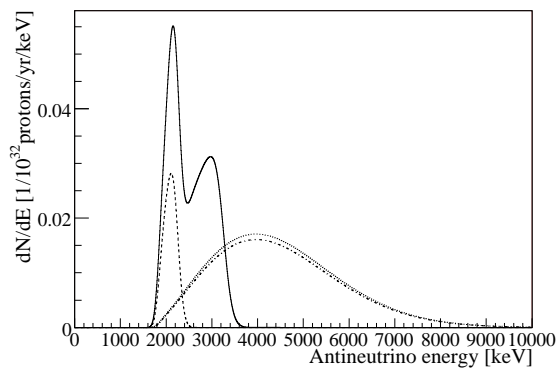}
\caption{The expected $\nuebar$ energy distribution with a detector energy resolution of $6\%/\sqrt{E{\rm[MeV]}}$ for the \A{U}{238} (solid) and \A{Th}{232} (dash) geoneutrinos, expected commercial reactor background (dot-dash), and the expected spectrum from a 6\,TW natural reactor at the Earth's core (dot).}
\label{fig:spectrumsObserved} 
\end{figure}

Figure~\ref{fig:spectrumsObserved} shows the expected spectra for geoneutrinos, commercial nuclear reactors, and a natural nuclear reactor at the Earth's core. The commercial reactor background is insignificant for the geoneutrino measurement. It is also small enough to allow an ultimate sensitivity of 1.3\,TW at 99\,\% CL, limited by the systematic uncertainty in the commercial reactor flux, for a nuclear reactor at the Earth's core. This assumes that the commercial reactor background can be obtained to 10\,\% accuracy, which should be possible based on the published electrical power and an averaged core nuclear cycle. If the isotopic fission rates of the reactors can be obtained the reactor background could be determined to $\sim~2\,\%$ accuracy\cite{Apollonio:2003, Boehm:2001,Eguchi:2003}.

\subsection{Radioactive backgrounds}
\label{sec:radioactiveBackgrounds}
The largest radioactive background in the recent KamLAND measurement was due to the reaction $^{13}{\rm C}(\alpha,{\rm n})$. The neutron produces two events, one as it loses energy, and the second when it captures on a proton. This mimics the $\nuebar$ events. The $\alpha$ in this reaction is a product of \A{Pb}{210} decay, which is itself a product from the decay of radon (Rn) gas present in the detector during construction. There is a plan to purify the KamLAND detector, reducing this background by a factor of one million, making this background negligible in the liquid scintillator.

The next most significant radioactive background is due to random correlations caused by radioactivity in the detector, mostly from U, Th, K, and Rn decays. The KamLAND experiment achieved U, Th, and \A{K}{40} concentrations in the scintillator of $6\times 10^{-16}\,{\rm g/g}$, $2\times 10^{-16}\,{\rm g/g}$, and $2\times 10^{-16}\,{\rm g/g}$, respectively. This resulted in negligible random coincidences due to radioactivity in the scintillator. However, radioactivity within the detector enclosure and surrounding rock, required a fiducial volume cut which reduced the effective detector mass.

Based on the results achieved with KamLAND, the purities required to perform this measurement for future experiments are clearly possible. However, the exact purity needed depends on the final detector design, discussed in Section~\ref{sec:detector}.

\subsection{Cosmic-ray backgrounds}
\label{sec:cosmicBackgrounds}
Cosmic-ray muons produce energetic neutrons and radioactive isotopes which can mimic the neutron inverse $\beta$ decay signature. The effect of energetic neutrons and cosmogenic radioactivity is reduced by vetoing the detector after a muon passes through. There is a small residual background due to muon vetoing inefficiency and backgrounds caused by muons that pass through the rock surrounding the detector without detection in the muon veto.

The recent KamLAND result\cite{Araki:2005} had a negligible background due to energetic neutrons and a background due to cosmogenic radioactivity of 0.6 per $10^{32}$ target proton yr. Because of the greater rock overburden at the 4850' level of the Homestake mine, the energetic neutron and cosmogenic backgrounds are expected to be $\sim 20$ times less than those at the KamLAND site\cite{Mei:2005}. The exact cosmic-ray background rates will depend on the detector material, layout, and veto efficiency, although it is expected that in almost any final design this will be negligible.

\section{The Detector}
\label{sec:detector}
In past experiments, both liquid scintillator and water Cherenkov detectors have been used to observe the positron and neutron produced in neutron inverse $\beta$ decay. Both techniques detect the photons emitted as charged particles move through the detector. The neutron is detected via the $\gamma$-ray emitted from its capture by a nucleus in the detecting material.

Water Cherenkov detectors produce a cone of light which allows the direction of the charged particle to be determined, and therefore allowing the direction of the $\nuebar$ to be inferred. The Cherenkov  photon yield is generally much less than the scintillation light yield, and consequently water Cherenkov detectors have poor sensitivity to events with energy less than $\sim 4\,{\rm MeV}$. A typical liquid scintillator detector easily observes $\nuebar$'s at the neutron inverse $\beta$ decay threshold. However, liquid scintillator detectors do not have very good directional information.

\subsection{Requirements}
\label{sec:detectorRequirements}
The following measurements and test should be performed by the proposed detector: measure the total geoneutrino rate, measure the ratio of \A{U}{238} to \A{Th}{232} in the Earth's interior, and test the hypothesis of a natural nuclear reactor at the Earth's core. Excluding the \A{U}{238} and \A{Th}{232} distribution in the Earth, the largest error in determining the expected geoneutrino rate is due to the uncertainty in the neutrino oscillation parameter $\sin^22\theta_{12}$ which is known to $\sim 6\,\%$. It is unlikely that the accuracy of this parameter will be determined to better than a few percent within the next decade. Therefore, it does not make sense to plan on measuring the geoneutrino rate to much better than 6\,\%. Assuming a 10\% error in the commercial nuclear reactor background, and only using the $\nuebar$ spectrum below 3.4\,MeV, to measure the total geoneutrino rate to $\sim 10\,\%$ the required exposure is estimated to be $\sim 2.3\times 10^{32}$ target proton yr. This does not include systematic errors, but these should be constrained to better than 10\,\%.

In determining the ratio of \A{U}{238} to \A{Th}{232} many errors cancel, therefore it should be possible to obtain a measurement to better than 10\,\% uncertainty. The Th/U ratio is currently estimated from meteorites to be between 3.7 and 4.1. To do this measurement, the $\nuebar$ energy spectrum could be split into two regions, one between 2.5\,MeV and 3.5\,MeV, which contains only \A{U}{238} $\nuebar$'s, and the other between 1.5\,MeV and 2.5\,MeV. An exposure of $\sim 20\times 10^{32}$ target proton yr is required to measure the ratio to 10\,\% accuracy. The uncertainty could be slightly reduced by a full spectral shape analysis.

Assuming a 10\% error in the commercial nuclear reactor background, and only using the $\nuebar$ spectrum above 4\,MeV, the required exposure to observe a 6TW georeactor at 3 sigma above zero is estimated to be $\sim 0.8\times 10^{32}$ target proton yr.

For a measurement of the total geoneutrino rate and an observation of a hypothetical georeactor, we need an exposure of about $2\times 10^{32}$ target proton yr. This could be achieved in approximately four years assuming a similar fiducial volume, $700\,{\rm m^3}$, and target proton density to KamLAND\cite{Eguchi:2003}. A detector much larger than this is not required, since this detector will already reach the sensitivity imposed by the uncertainty in the neutrino oscillation parameters. However, an accurate measurement of the \A{U}{238} to \A{Th}{232} ratio would require a larger detector, or a longer exposure time.

\subsection{Detector Design}
\label{sec:design}
There are two main types of large scintillator $\nuebar$ detectors: monolithic, such as the 1\,kton KamLAND detector\cite{Eguchi:2003}; and segmented, such as the 11\,ton Palo Verde detector\cite{Boehm:2001}. The advantage of a monolithic detector is reduced random coincidence backgrounds due to reduced support material, which is typically harder to purify than the scintillator. However, it would not be possible to build a KamLAND shaped detector at Homestake mine without further excavation since the detector is spherically symmetric. The advantage of a segmented detector is it could be constructed in sections above ground and transported below for assembly. Depending on the segment size, it could also be placed in one of the larger existing cavities.

In the KamLAND detector, the neutron produced in the neutron inverse $\beta$ decay is captured by a proton with a mean capture time of $\sim200\,\mu$s producing a 2.2\,MeV $\gamma$-ray. Gadolinium (Gd) was added to the Palo Verde detector scintillator in order to reduce backgrounds. Neutron capture by Gd produces $\gamma$-rays with a total energy of $\sim 8{\rm \,MeV}$, which is a higher energy than that produced by most radioactive backgrounds, greatly reducing the accidental background. Additionally, the Gd neutron capture cross-section is high, resulting in a mean neutron capture time of only $\sim27\,\mu$s in the Palo Verde detector, which would further decrease the accidentals due to the shorter correlation time window.

\section{Conclusion}
\label{sec:conclusion}
A measurement of geoneutrinos is an important step in constraining our understanding of the Earth's uranium and thorium distributions. The heat from the decay of these isotopes is the driving force for plate tectonics and earthquakes, and this is the only technique that allows us to directly observe these decays occurring at the inner depths of the Earth. The KamLAND experiment\cite{Araki:2005} has recently shown the viability of such a measurement; however, it was limited by backgrounds from nearby nuclear power plants. A similar experiment at the Homestake mine does not have the same problem with nearby nuclear power plants, and other backgrounds should be small or negligible, but depend on the final detector design. It is envisioned that a detector could be located in an existing cavity in the Homestake mine, such as the one used by the Davis experiment\cite{Davis:1968}.

\bibliography{homestakeGeoneutrinos} 

\end{document}